\documentclass[11pt,twoside]{article}

\usepackage{asp2006}

\begin{document}

\title{\hbox{\phantom{.}\hskip -40pt The~Elliptical--Spheroidal~and~Elliptical--Elliptical~Galaxy~Dichotomies}} 

\author{John Kormendy}   
\affil{Department of Astronomy, University of Texas at Austin, 1 University {\phantom{000}}Station C1400, 
       Austin, TX 78712-0259, USA}    

\begin{abstract} This paper summarizes Kormendy et al.~(2009, ApJS, in press, arXiv:0810.1681). 
We confirm that spheroidal galaxies have fundamental plane correlations 
that are almost perpendicular to those for bulges and ellipticals.  Spheroidals are not dwarf 
ellipticals.  They are structurally similar to late-type galaxies.  We suggest that they are 
defunct (``red~and~dead'') late-type galaxies transformed by a variety of gas removal processes. 
Minus spheroidals, ellipticals come in two varieties:~giant, non-rotating, boxy galaxies 
with cuspy cores and smaller, rotating, disky galaxies that lack cores.  We find a 
new feature of this ``E--E dichotomy'':~Coreless ellipticals have extra light at the 
center with respect to an inward extrapolation of the outer S\'ersic profile.  We suggest that 
extra light is made in starbursts that swamp core scouring in wet mergers.  In general, only giant, 
core ellipticals contain X-ray gas halos.  We suggest that they formed in mergers that were kept
dry by X-ray gas heated by active galactic nuclei. 
\pretolerance=15000  \tolerance=15000
\end{abstract}

\phantom{0}
\vskip -40pt
\phantom{0}

\section{High-Dynamic-Range Surface Photometry of E and Sph Galaxies}  

      Figure 1 shows examples of photometry of all known ellipticals in the 
Virgo cluster from Kormendy et al.~(2009; hereafter KFCB). The key to this paper is the 
accuracy and large dynamic range attained when data from many telescopes with different resolutions 
and field sizes are combined into composite profiles.  

\begin{figure}[h!]
\pretolerance=15000  \tolerance=15000 

\vskip 2.2truein

\includegraphics{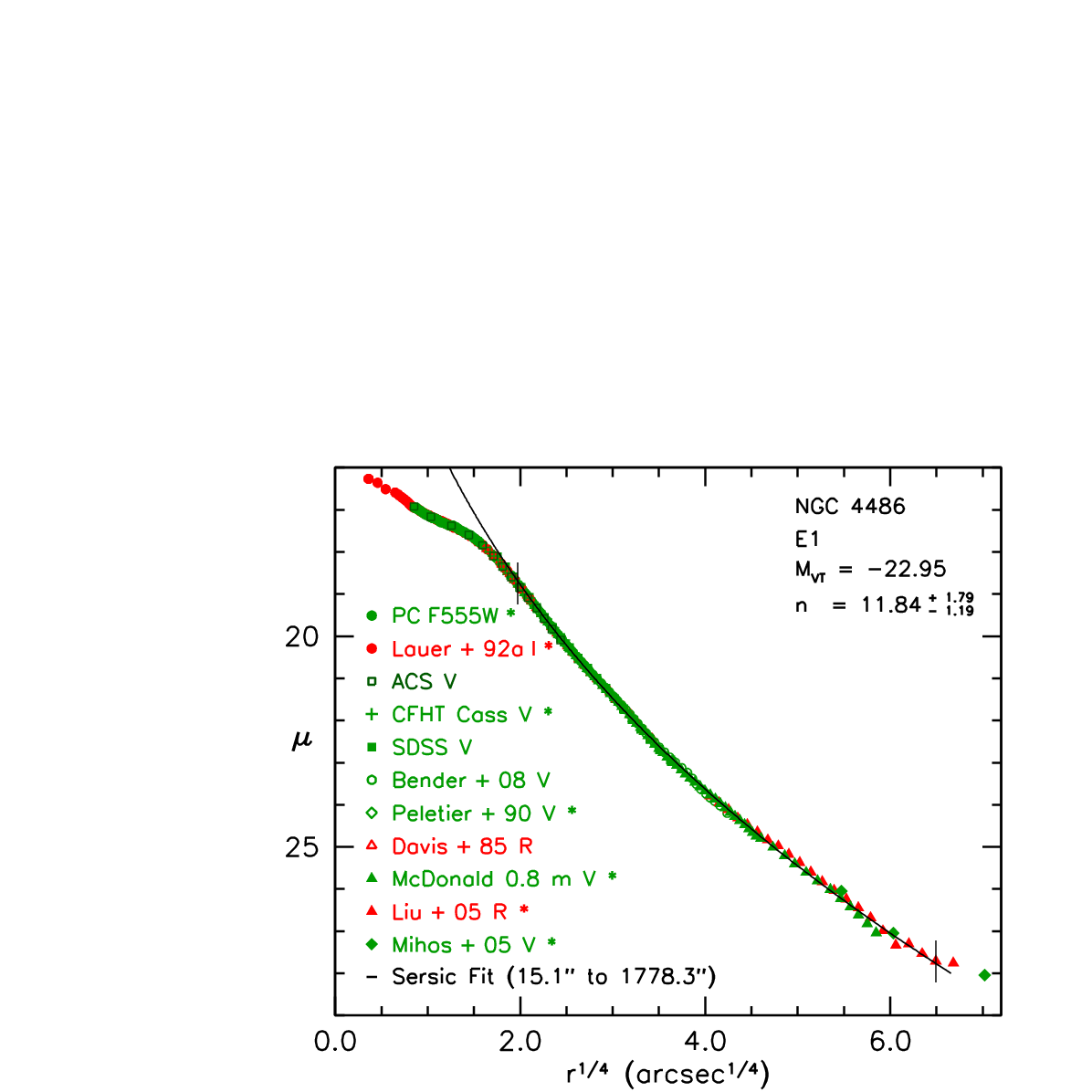}

\includegraphics{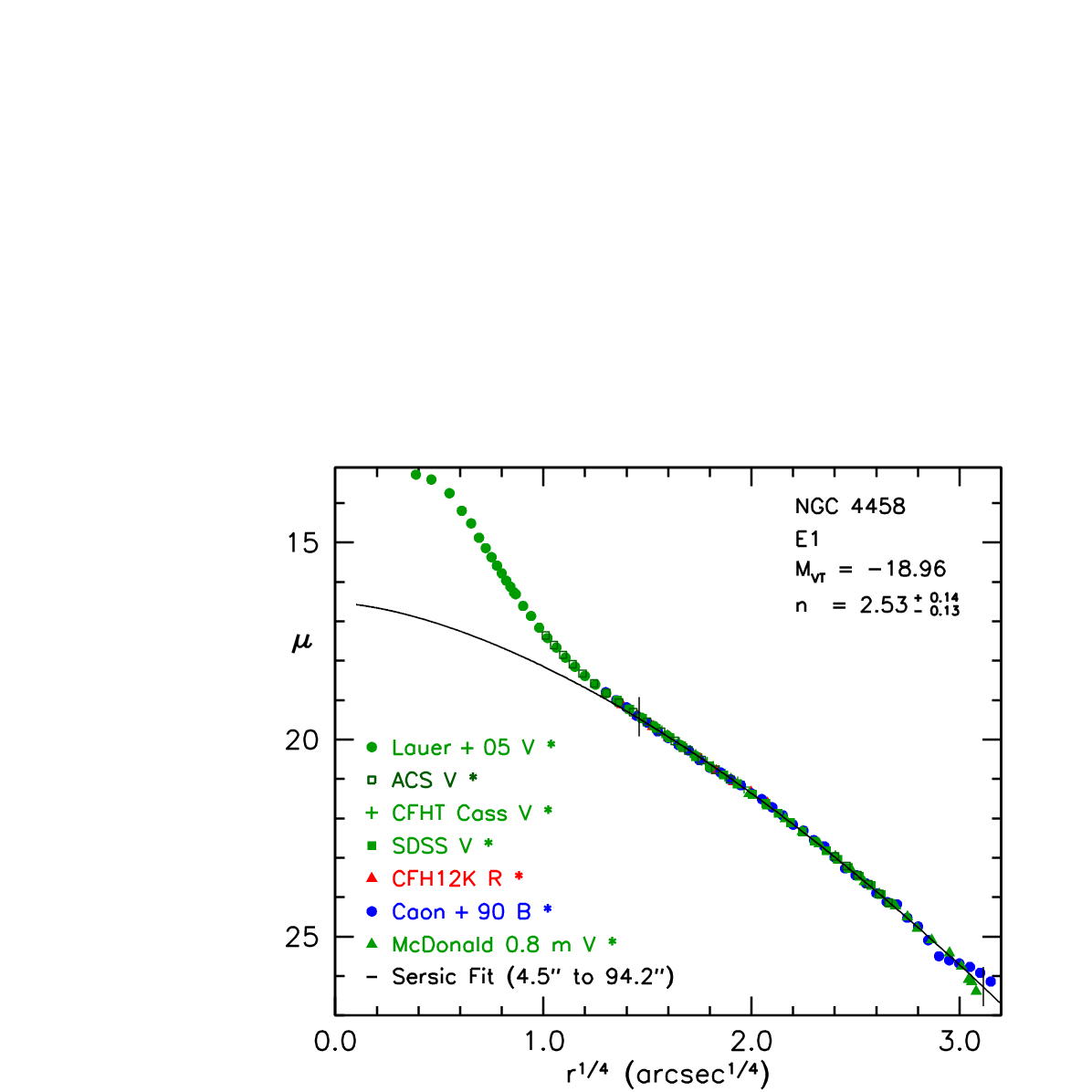}

\caption{\lineskip=-6pt \lineskiplimit=-6pt 
Major-axis brightness profiles of Virgo ellipticals from KFCB.  Surface brightness 
$\mu$ is in $V$ mag arcsec$^{-2}$.  The curve is a S\'ersic (1968) fit between the 
vertical dashes; the index $n$ and total galaxy absolute magnitude $M_{VT}$ are in the key. 
This figure illustrates the E--E dichotomy: NGC 4486 is a core galaxy with $n > 4$;
NGC 4458 is an extra light galaxy with $n \la 4$ (\S\thinspace3).
}
\end{figure}
\eject

\section{The E -- Sph Dichotomy} 

\pretolerance=15000  \tolerance=15000 

      Integrating the KFCB photometry leads to improved measurements of structural parameters. 
Figure 2 shows the resulting projections of the fundamental plane (``FP'') (Djorgovski \& Davis 
1987; Faber et al.~1987; Djorgovski et al.~1988).

\begin{figure}[hb!]
\plotfiddle{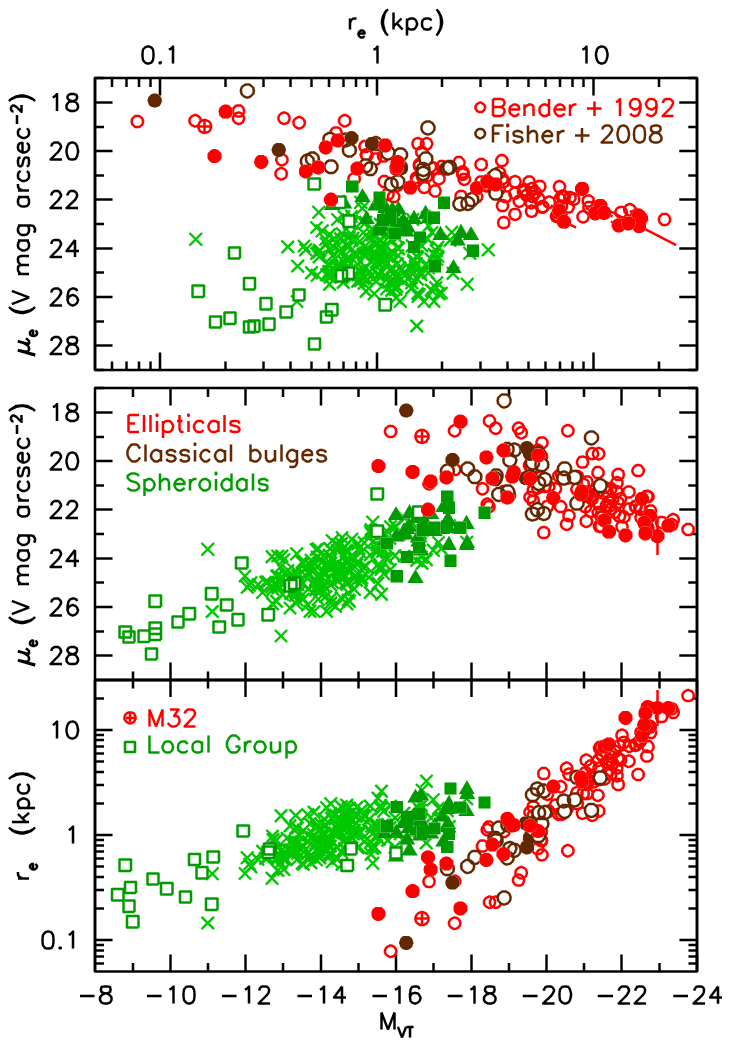}{5.3truein}{0}{82}{82}{-250}{-172}
\caption{\lineskip=0pt \lineskiplimit=0pt
Global parameter correlations for ellipticals (red), classical bulges (brown),
and spheroidals (green). Filled circles, filled squares, and M{\thinspace}32
are from KFCB.  Green triangles show all spheroidals from Ferrarese et al.~(2006)
that are not in KFCB.  Crosses show all spheroidals from Gavazzi et al.~(2005)
that are not in KFCB or in Ferrarese et al.~(2006).  The bottom panels show 
effective radius $r_e$ and surface brightness $\mu_e$ at the effective radius 
versus galaxy absolute magnitude.  The top panel shows $\mu_e$ vs.~$r_e$ 
(the~Kormendy 1977 relation, which shows the FP almost edge-on).  We confirm 
the \hbox{E{\thinspace}--{\thinspace}Sph} dichotomy found by Kormendy (1985,~1987).  
However, the separation between the E and Sph sequences is larger in near-central 
parameters, so KFCB use these to classify galaxies as elliptical or spheroidal.
\pretolerance=15000  \tolerance=15000 
}
\end{figure}

\eject

       KFCB and Figure 2 strongly confirm the reality of the E -- Sph dichotomy.  
Ellipticals and spheroidals form almost-perpendicular parameter sequences.
That E and Sph galaxies are physically different was presciently suggested 
by Wirth \& Gallagher (1984) based on a very few objects.  It was first demonstrated 
for substantial galaxy samples by Kormendy (1985, 1987).  

      The KFCB sample is augmented in Figure 2 with ellipticals from the FP study of 
Bender, Burstein, \& Faber (1992).  They, too, confirm 
the\thinspace\hbox{ E{\thinspace}--{\thinspace}Sph} dichotomy;
their sample substantially strengthens our results.  Bulges from 
Fisher \& Drory (2008) are also included.  We confirm that classical bulges are indistinguishable 
from ellipticals, as they should be if we understand correctly that they -- like ellipticals --
were formed in major mergers (Steinmetz \& Navarro 2002).~The same is not true of pseudobulges:
some of these are much less dense than classical bulges 
(Carollo 1999; Fisher \& Drory 2008; Kormendy \& Fisher 2008).  We believe they did not form via
mergers (Kormendy \& Kennicutt 2004).

      The existence of an E -- Sph dichotomy has been criticized by many authors, although
it is often visible in their correlation diagrams.  Early ambivalence is evident 
in Sandage, Binggeli, \& Tammann (1985).  They found continuous (but not monotonic) correlations 
involving $r_e$ and $\mu_e$, because their small Es were poorly resolved in
ground-based seeing.  This led them to believe that
Es and ``dEs'' (as they called spheroidals) are continuous.  On the other hand, they 
correctly noted that E and dE galaxies have different luminosity functions.  In fact,
``ellipticals'' and ``dwarf ellipticals'' overlap in luminosity!  The schizophrenic caption
of their Figure 7 reads: ``The luminosity functions for E and dE types suggesting that the two types, 
although part of a continuum of physical properties, form two families.''  Soon afterward, Binggeli 
\& Cameron (1991) saw the dichotomy in global parameters.  But Jerjen \& Binggeli (1997)
changed their minds: ``The [$n - M_B$] relation for Es and that for dEs smoothly and continuously
merge into each other, giving the impression of one global relation for dwarf and giant 
ellipticals.''  This observation, similarly interpreted by Graham \& Guzm\'an 
(2003), Gavazzi et al.~(2005), and Ferrarese et al.~(2006), is confirmed also by KFCB.
But the observation that $n$ is not sensitive to the difference between Es and Sphs does not mean
 that they are related. Another parameter that shows continuous E{\thinspace}--{\thinspace}Sph and even 
\hbox{E{\thinspace}--{\thinspace}S{\thinspace}--{\thinspace}Im{\thinspace}--{\thinspace}Sph} correlations
is metallicity, because self-enrichment depends
on potential well depth and not on structure.  To decide whether various galaxies are related, 
we need to look at \underbar{all} correlations.  Then Figure 2 shows that E and Sph galaxies are different.

      Other criticisms followed.  Graham \& Guzm\' an (2003) and Gavazzi et al.~(2005) argued
that core Es deviate from a continuous correlation beween Sphs and low-luminosity Es because of 
the light that is missing in cores.  This is unrealistic: only 0{\thinspace}--{\thinspace}2\thinspace\% 
of the galaxy light is missing; such small amounts have no effect on measurements of global
parameters. Moreover, the Sph sequence approaches the E sequence near its middle, 
not near its faint end.  Similarly, it is unrealistic to argue (Binggeli 1994; Graham \& Guzm\' an 2003; 
Gavazzi et al.~2005; Ferrarese et al.~2006) that ``the striking dichotomy observed by Kormendy (1985) 
could partly be due to the lack, in Kormendy's sample, of galaxies in the $-20~{\rm mag} < M_B < 
-18~{\rm mag}$ range, corresponding precisely to the transition region between the two families.''
Kormendy (1985, 1987) showed that the families strongly diverge in the magnitude range where 
they overlap.  Also, KFCB and Figure 2 here suffer from no lack of galaxies at overlap  magnitudes.  
Finally, the Sph galaxies in KFCB are biased in favor of those that are most like ellipticals,
because we did not know which galaxies were E and which were Sph -- or whether these could 
be distinguished -- until the photometry was completed. 

      Thus the E -- Sph dichotomy is a robust result, visible in central parameters
(KFCB Figure 34), in global parameters (KFCB Figures 37, 38, 41; Figure 2 here) and in 
the surface brightness profiles (KFCB Figures 35, 36, 39).

      Figure 2 shows that lower-luminosity Es are monotonically higher in density, whereas
lower-luminosity Sphs are monotonically lower in density.  Ellipticals define a sequence of increasing 
dissipation in lower-mass mergers (Kormendy 1989).  It is reproduced by
merger simulations that include gas and star formation (Robertson et al.~2006; Hopkins et al.~2008c, d).  
In contrast, Kormendy (1985, 1987) showed that the parameter correlations of Sph galaxies are similar 
to those of dwarf spiral and irregular galaxies.   The above papers and KFCB suggest that spheroidals 
are defunct late-type galaxies transformed by internal processes such as supernova-driven gas ejection 
(Dekel \& Silk 1986) and environmental processes such as galaxy harassment (Moore et al.~1996, 1998) and 
ram-pressure gas stripping (Chung et al.~2007).  Smaller Sph\thinspace$+${\thinspace}S\thinspace$+${\thinspace}Im 
galaxies form a sequence of decreasing baryon retention caused by the shallower gravitational 
potential wells of tinier galaxies (Dekel \& Silk 1986).

\vskip -20pt

\centerline{\null}

\section{The E -- E Dichotomy} 

\pretolerance=15000  \tolerance=15000 

      Minus spheroidals, ellipticals and classical bulges of disk galaxies 
(red~and~brown points in Figure 2) form a homogeneous collection of objects that are 
consistent with our paradigm of galaxy formation.  We believe that both formed in 
galaxy mergers that are an inevitable consequence of the hierarchical clustering that makes all 
structure in the Universe (Toomre~1977; White \& Rees 1978; Steinmetz \& Navarro 2002).  We see 
mergers in progress, often with gas dissipation and bursts of star formation (Schweizer 1990).  
Most spectacular are ultraluminous infrared galaxies (ULIRGs) (e.{\thinspace}g., Joseph \& Wright 
1985; Sanders et al.~1988).  Their remnants are quantitatively consistent with the properties of
ellipticals.  What do we still need to learn?

      Research now focuses on the exact sequence of events that made ellipticals.  
One complication is that star formation and galaxy assembly via mergers can be very 
disconnected.  Which galaxies are remnants of gas-free (``dry'') mergers and which are remnants 
of gas-rich (``wet'') mergers?  Also, close correlations of bulges and ellipticals 
with their central black holes (BHs) imply that BHs and their hosts evolved together (Ho 2004).  
BH growth may inject so much energy into protogalactic gas that it becomes a dominant force in 
galaxy formation (Silk \& Rees 1998; Hopkins et al.~2006).  How do BHs affect galaxy formation?  
Third, much of the emphasis in understanding galaxy formation
has shifted from explaining structure to explaining star formation histories in the context of the 
Sloan Digital Sky Survey observations of the color-magnitude diagram (e.{\thinspace}g., Strateva et 
al.~2001) in which galaxies divide themselves into a ``red sequence'' of passively evolving objects 
and a ``blue cloud'' of galaxies that actively form stars.  Ellipticals dominate the bright
part of the red sequence.  How did they evolve from fainter red-sequence and blue-cloud galaxies? 
What observed properties of ellipticals provide clues to an understanding of the above issues?

      The most fundamental such properties consist of the observations that elliptical galaxies come 
in two distinct varieties.  In the  following list of properties, italics highlight new aspects of 
the dichotomy found in KFCB or known aspects for which the observational evidence is strengthened in KFCB.

      \underbar{Giant ellipticals} ($M_V \la -21.5 \pm 1$ for $H_0 = 72$ km s$^{-1}$ Mpc$^{-1}$)
generally \hfill\break
(1) have cores,  i.{\thinspace}e., central missing light with respect to the outer profile (Fig.~1);\hfill\break
(2) rotate slowly, so rotation is of little importance dynamically; hence \hfill\break
(3) are anisotropic and modestly triaxial; \hfill\break
(4) are less flattened (ellipticity $\sim${\thinspace}0.15) than smaller ellipticals; \hfill\break
(5) have boxy-distorted isophotes; \hfill\break
(6) {\it have S\'ersic function outer profiles with $n > 4$} (Fig.~1); \hfill\break
(7) {\it mostly are made of very old stars that are enhanced in $\alpha$ elements} (Fig.~3). \hfill\break
(8) often contain strong radio sources, and \hfill\break
(9) {\it contain X-ray-emitting gas, more of it in more luminous galaxies} (Fig.~4).

      \underbar{Normal and dwarf ellipticals} ($M_V \ga -21.5$) generally \hfill\break
(1) are coreless -- {\it have central extra light with respect to the outer profile} (Fig.~1);\hfill\break
(2) rotate rapidly, so rotation is dynamically important to their structure;  \hfill\break
(3) are nearly isotropic and oblate spheroidal, albeit with small axial dispersions; \hfill\break
(4) are flatter than giant ellipticals (ellipticity $\sim${\thinspace}0.3); \hfill\break
(5) have disky-distorted isophotes; \hfill\break
(6) {\it have S\'ersic function outer profiles with $n \la 4$} (Fig.~1); Fig.~3: \hfill\break
(7) {\it are made of (still old but) younger stars with little $\alpha$-element enhancement}; 
(8) rarely contain strong radio sources, and \hfill\break
(9) {\it rarely contain X-ray-emitting gas} (Fig.~4). 

      These results are etablished in many papers (e.{\thinspace}g.,
Davies et al.~1983;
Bender 1988;
Bender et al.~1989; 
Nieto et al.~1991;
Kormendy et al.~1994; 
Lauer et al.~1995, 2005, 2007;
Kormendy \& Bender 1996;
Tremblay \& Merritt 1996;
Gebhardt et al.~1996; 
Faber et~al.~1997; 
Thomas et al.~2005;
Emsellem et al.~2007;
Cappellari et al.~2007).
A few ellipticals are exceptions to $\geq$ one of (1) -- (9).

     How did the E{\thinspace}--{\thinspace}E dichotomy arise?  The ``smoking gun'' for an explanation
is a new aspect of the dichotomy found in KFCB and illustrated here~in~Figure~1.  Coreless galaxies 
do not have featureless, central ``power law profiles''.  Rather, all coreless galaxies in the KFCB
sample show a new structural component, i.{\thinspace}e., extra light near the center above the 
inward extrapolation of the outer S\'ersic profile (e.{\thinspace}g,, M{\thinspace}32: Kormendy 1999;
NGC 4458: Fig.~1, right).~Kormendy~(1999) suggested that this extra light is the signature of 
starbursts produced in dissipative mergers as predicted by Mihos \& Hernquist's (1994) simulations.
Similar extra light components are seen in all coreless ellipticals in KFCB.  Like Faber et al.~(1997, 2007),
KFCB suggest that the origin of the E -- E dichotomy is that core Es formed in dry mergers whereas 
coreless Es formed in wet mergers.  Simulations of dry and wet mergers now reproduce the structural 
properties of core and extra light ellipticals in beautiful detail (Hopkins et al.~2008a, b).

      When did the E{\thinspace}--{\thinspace}E dichotomy arise?  We cannot answer yet,
but Figure~3 provides constraints.  It shows observation (7) that core ellipticals
mostly are made of old stars that essentially always are enhanced in $\alpha$ elements; in
contrast, coreless ellipticals are made of younger stars with more nearly solar compositions. 
This means (Thomas et al.~2005) that the stars that now are in core Es formed in the first few billion
years of the Universe and over a short period of $\la 1$ Gyr,

\centerline{\null}

\vskip 2.87truein

\begin{figure}[ht!]

\includegraphics{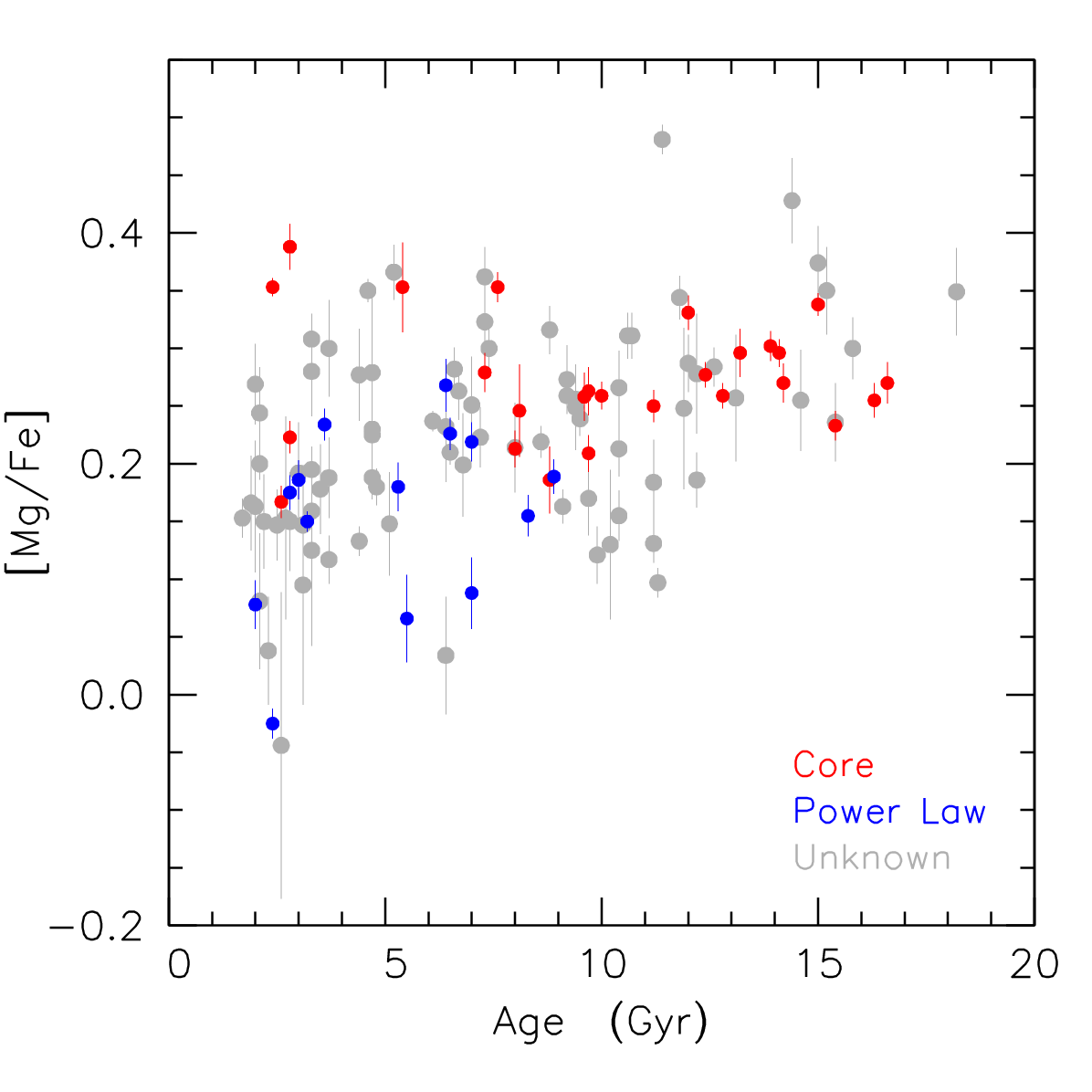}

\caption{\lineskip=0pt \lineskiplimit=0pt
Alpha element overabundance ($\log$ solar units) versus relative age of the stellar
population (data from Thomas et al.~2005).  Red and blue points denote core and power 
law ellipticals.  This figure is from KFCB.
}
\end{figure}

\noindent so quickly that Type I supernovae did not have time to dilute with Fe the 
\hbox{$\alpha$-enriched} gas recycled by Type II supernovae. However, this 
does not mean that core ellipticals were made at the same time as their stars.~Mass assembly 
via dry mergers -- as required to explain their structure -- could have happened at 
any time after star formation stopped.  Our problem is to explain how star formation was 
quenched so quickly and not allowed to recur.  In contrast, coreless ellipticals have younger, 
less-$\alpha$-enhanced stellar populations.  They are consistent with a simple picture in 
which a series of wet mergers with accompanying starbursts formed their stellar populations 
and assembled the galaxies more-or-less simultaneously over the past 9 billion years.  
Faber et al.~(2007) discuss these issues in detail. A corollary is that the progenitors of
coreless ellipticals likely were more similar to present-day galaxies than were the
progenitors of core ellipticals.  The latter may have been different from all galaxies seen today.

          Why did the E{\thinspace}--{\thinspace}E dichotomy arise?~The key observations are:~(8) 
core-boxy ellipticals often are radio-loud whereas coreless-disky ellipticals are not, and 
(9) core-boxy ellipticals contain \hbox{X-ray} gas whereas coreless-disky ellipticals do not
(Bender et al.~1989).  Figure 4 brings result (9) \hbox{up-to-date.}  KFCB suggest that the
hot gas keeps dry mergers dry and protects giant ellipticals from late star formation.  This
is the operational solution to the above ``maintenance problem''.  However, the trick is to keep the 
gas hot.  It is well known that X-ray gas cooling times are short.  KFCB review evidence that 
the main heating mechanism may be energy feedback from accreting BHs (the AGNs of observation 8);
these may also have quenched star formation after $\sim 1$ Gyr.  Many details of this picture require work 
(Cattaneo et al.~2009).  Cosmological gas infall is an additional heating mechanism 
(Dekel \& Birnboim 2006).  Nevertheless, Figure 4 provides a crucial connection 
between \hbox{X-ray} gas, AGN physics, and the E -- E dichotomy.

\vfill\eject

\centerline{\null}

\vskip 2.7truein

\begin{figure}[ht!]

\includegraphics{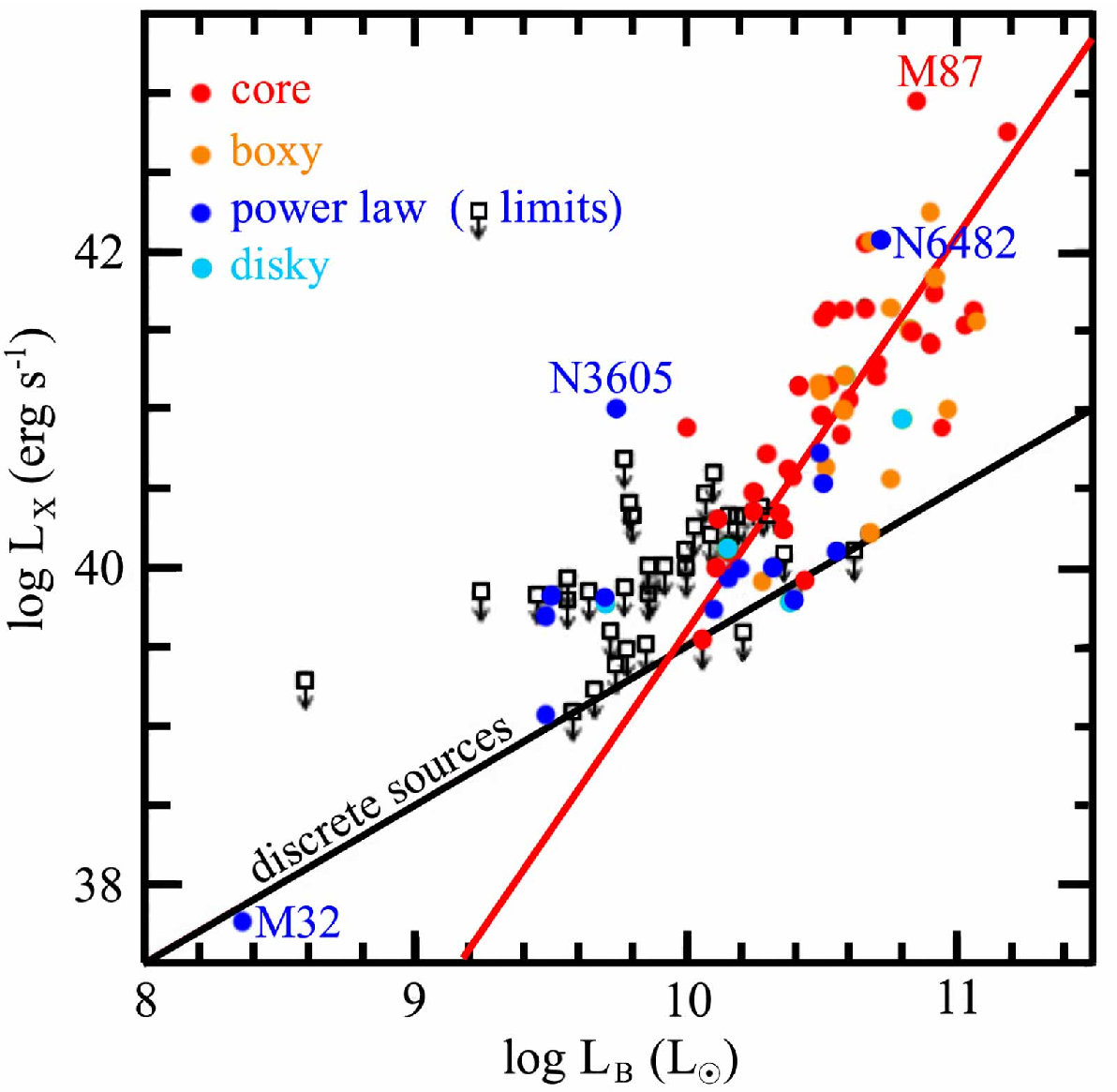}

\caption{\lineskip=0pt \lineskiplimit=0pt
Observed X-ray emission versus galaxy luminosity (KFCB; adapted from Fig.~9 of Ellis \& 
O'Sullivan 2006).  Detections are color-coded according to the E -- E dichotomy (see the key).  
The black line is an estimate of the contribution from discrete sources.  The 
red line is a bisector fit to the core-boxy points.  Core and coreless (``power law'') Es 
overlap in luminosity, but most core galaxies do and most coreless galaxies do not contain 
hot gas. 
}
\end{figure}

      ``\underbar{Bottom line}:'' In essence, only giant, core ellipticals and their progenitors 
are massive enough to contain hot gas that helps to engineer the E{\thinspace}--{\thinspace}E dichotomy.

\acknowledgements I thank Ralf Bender, Mark Cornell, and David Fisher for fruitful
collaboration.~This work was supported by NSF grant AST-0607490.

\centerline{\null}
\vskip -31pt
\centerline{\null}


\begin{thebibliography}{}

\bibitem[]{} Bender, R.~1988, A\&A, 193, L7 

\bibitem[]{} Bender, R., Burstein, D., \& Faber, S.~M.~1992, ApJ, 399, 462

\bibitem[]{} Bender, R., et al.~1989, A\&A, 217, 35

\bibitem[]{} Binggeli, B.~1994, in ESO/OHP Workshop on Dwarf Galaxies, ed.~G.~Meylan \&
             P.~Prugniel (Garching: ESO), 13

\bibitem[]{} Binggeli, B., \& Cameron, L.~M.~1991, A\&A, 252, 27

\bibitem[]{} Cappellari, M., et al.~2007, MNRAS, 379, 418 

\bibitem[]{} Carollo, C.~M.~1999, ApJ, 523, 566

\bibitem[]{} Cattaneo, A., et al.~2009, Nature, submitted

\bibitem[]{} Chung, A.~et al.~2007, ApJ, 659, L115

\bibitem[]{} Davies, R.~L., et al.~1983, ApJ, 266, 41

\bibitem[]{} Dekel, A., \& Birnboim, Y.~2006, MNRAS, 368, 2

\bibitem[]{} Dekel, A., \& Silk, J.~1986, ApJ, 303, 39

\bibitem[]{} Djorgovski, S., \& Davis, M.~1987, ApJ, 313, 59

\bibitem[]{} Djorgovski, S., de Carvalho, R., \& Han, M.-S.~1988, in The Extragalactic 
             Distance Scale, ed.~S.~van den Bergh \& C.~J.~Pritchet (San Francisco: ASP), 329

\bibitem[]{} Ellis, S.~C., \& O'Sullivan, E.~2006, MNRAS, 367, 627

\bibitem[]{} Emsellem, E., et al.~2007, MNRAS, 379, 401 

\bibitem[]{} Faber, S.~M., et al.~1987, in Nearly Normal Galaxies: From the Planck
             Time to the Present, ed. S.~M.~Faber (New York: Springer), 175 

\bibitem[]{} Faber, S.~M., et al.~1997, AJ, 114, 1771 

\bibitem[]{} Faber, S.~M., et al.~2007, AJ, 665, 265

\bibitem[]{} Ferrarese, L., et al.~2006, ApJS, 164, 334 

\bibitem[]{} Fisher, D.~B., \& Drory, N.~2008, AJ, 136, 773

\bibitem[]{} Gavazzi, G., et al.~2005, A\&A, 430, 411

\bibitem[]{} Gebhardt, K., et al.~1996, AJ, 112, 105 

\bibitem[]{} Graham, A.~W., \& Guzm\'an, R.~2003, AJ, 125, 2936

\bibitem[]{} Ho, L.~C., Ed.~2004, Carnegie Observatories Astrophysics Series, Vol.~1,
             Coevolution of Black Holes and Galaxies (Cambridge: Cambridge University Press)

\bibitem[]{} Hopkins, P.~F., et al.~2006, ApJS, 163, 1 

\bibitem[]{} Hopkins, P.~F., et al.~2008a, ApJ, in press (arXiv:0805.3533)  

\bibitem[]{} Hopkins, P.~F., et al.~2008b, ApJ, submitted (arXiv:0806.2325)  

\bibitem[]{} Hopkins, P.~F., Cox, T.~J., \& Hernquist, L.~2008c, ApJ, in press (arXiv:0806.3974) 

\bibitem[]{} Hopkins, P.~F., et al.~2008d, ApJ, in press (arXiv:0807.2868)  

\bibitem[]{} Jerjen, H., \& Binggeli, B.~1997, in The Second Stromlo Symposium:
             The Nature of Elliptical Galaxies, ed.~M. Arnaboldi, et al.~(San Francisco:
             ASP), 239

\bibitem[]{} Joseph, R.~D., \& Wright, G.~S.~1985, MNRAS, 214, 87

\bibitem[]{} Kormendy, J.~1977, ApJ, 218, 333

\bibitem[]{} Kormendy, J.~1985, ApJ, 295, 73

\bibitem[]{} Kormendy, J.~1987, in Nearly Normal Galaxies: From the Planck Time
             to the Present, ed.~S.~M.~Faber (New York:~Springer), 163

\bibitem[]{} Kormendy, J.~1989, ApJ, 342, L63

\bibitem[]{} Kormendy, J.~1999, in Galaxy Dynamics: A Rutgers Symposium, ed.~D.~Merritt,
             J.~A.~Sellwood, \& M.~Valluri (San Francisco: ASP), 124

\bibitem[]{} Kormendy, J., \& Bender, R.~1996, ApJ, 464, L119

\bibitem[]{} Kormendy, J., \& Fisher, D.~B.~2008, in Formation and 
             Evolution of Galaxy Disks, ed.~J.~G.~Funes, S.{\thinspace}J.~\& E.~M.~Corsini 
             (San Francisco: ASP), 297 

\bibitem[]{} Kormendy, J., Fisher, D.~B., Cornell, M.~E., \& Bender, R.~2009, ApJS, in press (KFCB)
             (arXiv:0810.1681) 

\bibitem[]{} Kormendy, J., \& Kennicutt, R.~C.~2004, ARA\&A, 42, 603 

\bibitem[]{}  Kormendy, J., et al.~1994, in ESO/OHP Workshop on Dwarf Galaxies, 
              ed.~G. Meylan \& P.~Prugniel (Garching: ESO), 147

\bibitem[]{} Lauer, T.~R., et al.~1995, AJ, 110, 2622 

\bibitem[]{} Lauer, T.~R., et al.~2005, AJ, 129, 2138

\bibitem[]{} Lauer, T.~R., et al.~2007, ApJ, 664, 226 

\bibitem[]{} Mihos, J.~C., \& Hernquist, L.~1994, ApJ, 437, L47

\bibitem[]{} Moore, B., Lake, G., \& Katz, N.~1998, ApJ, 495, 139  

\bibitem[]{} Moore, B., et al.~1996, Nature, 379, 613  

\bibitem[]{} Nieto, J.-L., Bender, R., \& Surma, P.~1991, A\&A, 244, L37 

\bibitem[]{} Robertson, B., et al.~2006, ApJ, 641, 21

\bibitem[]{} Sandage, A., Binggeli, B., \& Tammann, G.~A.~1985, in ESO Workshop on the
             Virgo Cluster of Galaxies, ed. O.-G.~Richter \& B.~Binggeli (Garching: ESO), 239

\bibitem[]{} Sanders, D.~B., et al.~1988, ApJ, 325, 74

\bibitem[]{} Schweizer, F.~1990, in Dynamics and Interactions of Galaxies, 
             ed.~R.~Wielen (New York: Springer), 60

\bibitem[]{} S\'ersic, J.~L.~1968, Atlas de Galaxias Australes (Cordoba: 
           Obs.~Astr., Univ.~de Cordoba)

\bibitem[]{} Silk, J., \& Rees, M.~J.~1998, A\&A, 331, L1

\bibitem[]{} Steinmetz, M., \& Navarro, J.~F.~2002, NewA, 7, 155

\bibitem[]{} Strateva, I., et al.~2001, AJ, 122, 1861 

\bibitem[]{} Thomas, D., et al.~2005, ApJ, 621, 673

\bibitem[]{} Toomre, A.~1977, in The Evolution of Galaxies and Stellar 
             Populations, ed.~B.~M. Tinsley \& R.~B.~Larson (New Haven:
             Yale University Observatory), 401

\bibitem[]{}  Tremblay, B., \& Merritt, D.~1996, AJ, 111, 2243

\bibitem[]{} White, S.~D.~M., \& Rees, M.~J.~1978, MNRAS, 183, 341

\bibitem[]{} Wirth, A., \& Gallagher, J.~S.~1984, ApJ, 282, 85

\end{thebibliography}
\end{document}